\documentclass[nologo,11pt,a4paper]{ETHpaper}
\usepackage[square,numbers,sort&compress]{natbib}

\usepackage{amssymb,amsmath, epsfig,  graphicx}
\usepackage{listings}
\usepackage{hyperref}

\begin{document}
\title{\mbox{A complementary view} on the \mbox{growth of directory trees}}

\titlealternative{A complementary view on the growth of directory trees
}

\author{\mbox{Markus M. Geipel$^\star$\footnote{Corresponding author:
      \url{mgeipel@ethz.ch}}, Claudio J. Tessone$^\star$}, Frank Schweitzer$^\star$}

\authoralternative{M. M. Geipel, C. J. Tessone, F. Schweitzer}

 \address{$^\star$Chair of Systems Design, ETH  Zurich, Kreuzplatz 5, 8032 Zurich, Switzerland}


 \www{\url{http://www.sg.ethz.ch}}

\makeframing
\maketitle

\begin{abstract}
Trees are a special sub-class of networks with unique
  properties, such as the level distribution which has often been
  overlooked. We analyse a general tree growth model proposed by Klemm
  {\em et. al.} (2005) to explain the growth of user-generated directory
  structures in computers. The model has a single parameter $q$ which
  interpolates between preferential attachment and random growth. Our
  analysis results in three contributions: First, we propose a more
  efficient estimation method for $q$ based on the degree distribution,
  which is one specific representation of the model. Next, we introduce
  the concept of a level distribution and analytically solve the model
  for this representation. This allows for an alternative and independent
  measure of $q$. We argue that, to capture
   real growth processes, the $q$ estimations from the degree and the
   level distributions should coincide. Thus, we finally apply both
   representations to validate the 
  model with synthetically generated tree structures, as well as with
  collected data of user directories. In the case of real directory
  structures, we show that $q$ measured from the level distribution are
  incompatible with $q$ measured from the degree distribution. In
  contrast to this, we find perfect agreement in the case of
  simulated data. Thus, we
  conclude that the model is an incomplete description of the growth of real
  directory structures as it fails to reproduce the level distribution.
  This insight can be generalised to point out the importance of the
  level distribution for modeling tree growth.
\end{abstract}

\section{Introduction}
\label{sec:introduction_epjb}

Tree structures are pervasive in natural systems as well as in artificial
ones \citep{caldarelli05}. For example, in geology, river networks are a
paradigmatic example \cite{rodrigueziturbe97}. Moreover, trees also
appear in biology, for example in the vascular systems of animals and
plants \cite{zamir99,weibel91}. Recently, it was shown that these
transport systems exhibit universal scaling properties, which only depend
on the dimensionality of the space they are embedded in \cite{banavar99}.
Apart from that, trees are fundamental in computer models of plant
growth, also called Lin\-den\-mayer-systems, \cite{prusinkiewicz90}.

Trees are not only pervasive in nature but also in the way humans
structure knowledge and information: Different species have been
historically classified based on trees where each node represents one
species. First, through the linnaean taxonomic classification, where the
complete hierarchy is known as the {\em tree of life}
\cite{cracraft04}. Later through more evolved techniques, such as
cladorams \cite{dupuis84}, and (more recently) phylogenetic trees which
have helped to understand the diversification patterns at increasing
resolution \cite{rokas06}. Interestingly, these phylogenetic trees show
an outstanding invariance when seen at different scales, ranging from
inter- to intra-species ones \cite{tessone07c,hernandez-garcia07}.
Another example of trees is the categorisation of entries in
Wikipedia. Even though Wikipedia is non-hierarchically organised, the
categorisation forms an emergent tree structure \cite{muchnik07}.

Likewise, trees are dominant in computer systems. They are a fundamental
concept of algorithms: data compression, sorting, searching and analysis
of recursion are all tied to often highly sophisticated hierarchical
structures \cite{huffman52,knuth97,goodrich02}.  This also applies to one
of the most obvious tree in everyday work life: the directory structure
in our computers. The first popular fully hierarchical file system was
introduced with the UNIX operating system. Despite new non-hierarchical
organisation paradigms such as tagging \cite{golder05} or relational data
bases \cite{codd70}, the hierarchical organisation in directories remains
the indispensable basis of data storage in all modern computer systems. A
model to describe the growth of these directory trees has been proposed
in \cite{klemm05,klemm06}.

From a formal point of view, trees are a special sub-class of
networks. For example, in the network growth model by Krapivsky {\em et
  al.} \cite{krapivsky00}, if the number of added edges per time unit is
one, the resulting network is a tree.  Furthermore, each weighted network
can easily by reduced to a {\em minimum spanning tree}.  This method was
for example used to describe the backbones of complex networks
\cite{garlaschelli03,garlaschelli05}. The fact that trees are a sub-class
of networks, however, should not lead to the misconception that they are
trivial. Indeed they often show a high degree of complexity and offer a
set of unique properties, not existent in general networks. For example,
many 
existing tree structures exhibit scaling laws in the sub-tree size or
branch size distribution, named {\em allometric scaling}
\cite{caldarelli05}. Furthermore, in trees there is a special node, the
{\em root}, from which the tree grows. Thus, all trees also possess a
level distribution as a characteristic property.  Given these significant
differences between networks and trees and their remarkable features,
such allometric scaling and level distribution, the tools developed for
complex networks are not sufficient to capture the idiosyncratic
properties of trees.

Notwithstanding this insight, trees are all to often just treated as
simplified networks.  The aim of this paper is to fill this gap.
We focus on the tree growth model presented in \cite{klemm05}. Although
introduced as a model to explain the growth of computer directories, this
model constitutes a very general and straight-forward approach to the
growith hierarchical structures. As the main idea, it interpolates two
fundamental growth mechanisms: random growth and preferential attachment.
In this paper we complement the results on this general model in several
ways: We show that, when rewritten in terms of the level distribution,
the equations describing the growth of the tree can be solved and easily
validated against the data. Moreover, we introduce an alternative method
to estimate the parameters of the model based on the degree
distribution. We find that both methods allow us to obtain unbiased,
independent estimations of the relevant model parameters. Finally we
contrast the parameter estimation for computer simulated data of the
model with real world data. We confirm that the model presented in
\cite{klemm05} reproduces the properties of the degree distribution of
user generated directories, but we find that it falls short in
reproducing the corresponding level distribution.

The paper is organised as follows. In Section \ref{sec:model_epjb} we
review the stochastic model of Refs.~\cite{klemm05,klemm06} and the main
results therein. In Section \ref{sec:results_epjb} we solve two
complementary representations of the stochastic model: one written in
terms of the degree distribution, and another in terms of the level
distribution. Section \ref{sec:comparison_epjb} shows the comparison
between the estimation of the relevant parameters with simulations and
data gathered from different computer pools. The closing Section
\ref{sec:conclusion_epjb} presents the final summary and discusses the
main results.

\section{Model}
\label{sec:model_epjb}

The model introduced in Ref.~\cite{klemm05} interpolates between two
growth processes: one based on preferential attachment, and the
other based on random attachment. Initially, at $t=1$, there is one node:
the ``root'' node.  Then, at every time step $t$, a node is added to the
tree by one of two different processes: (i) with probability $q$, the
node is added following a preferential attachment rule: the larger the
in-degree ($k-1$) of a node, the more probable the new node is linked to
it. (ii) otherwise, with probability $1-q$, the node is added at random
to one of the existing ones.  Thus, at time $t$ the network size is
$N=t$. Throughout this Paper, we will use $N$ and $t$ interchangeably
depending on the context.

The probability of adding a node to an existing one with degree $k$ is
defined by the following equation:
\begin{equation}
  \label{eq:model}
  \Pi(k)=q\frac{k-1}{N}+(1-q)\frac{1}{N}.
\end{equation}
The normalisation of the second term (on the right-hand side) is
straight-forward: each node is equally likely to be chosen at random;
thus it is divided by $N$, the number of nodes in the system. The
normalisation of the first term deserves a brief explanation. First, it
is assumed that edges in the tree are directed from child to parent. Each
node has thus an out-degree of $1$. The in-degree is consequently
$k-1$. The proper normalisation would be $N-2$ as in a tree the sum of
all degrees equals $2(N-1)$. We assume however that the root node has an
initial degree of $2$, otherwise in the case of $q=1$ and time $t=1$,
$\Pi(k)$ for the only existing node, root, would be zero. For this
reason, also the $q$ term is normalised with $N$.

The authors of Ref~\cite{klemm05} verified this model against real
directory data in two ways. First, by a comparison of the allometric
scaling defined by the model and the one found in the data. The authors
showed that the model matches the data in this respect.  In the second
test, the authors calculated from the data the second, third, and fourth
moment of the degree distribution as well as the average distance between
nodes. For each of these four observed variables, the most probable value
of $q$ was then estimated by extensive computer simulation of the model,
rejecting/accepting randomly drawn values of $q$ via a Monte Carlo
method. The authors found an excellent agreement between these values of
$q$ estimated independently.

Apart from these tests, Ref.~\cite{klemm05} shows that the degree
distributions of the directory trees exhibit a non-universal exponent
while the scaling exponent of the distribution of branch sizes
(i.e.~sub-tree size distribution) is a power law with a universal
exponent which equals 2. In Ref.~\cite{klemm06} these findings were
complemented: In directory structures, the average distance to the root
increases logarithmically with system size and the exponent of the
allometric scaling is in all the cases close to 1.

\section{Analysis}
\label{sec:results_epjb}
\subsection{Degree distribution}
\label{sec:degree-distribution_epjb}

In this section we present the results of our analysis of the model
defined in equation~(\ref{eq:model}). The first part is dedicated to the
\emph{degree distribution} generated by the model, while
Section~\ref{sec:level-distr-nodes_epjb} addresses the \emph{level
  distribution}.

Just like networks, trees have a certain degree distribution which
depends on their growth process.  To analyse this, we first write down
the exact discrete equations for the evolution of this distribution over
time.  Next, we present closed forms for the recursive solution and
analyse their validity.  Finally, we analyse how far concrete
realisations of trees grown based on the model defined in Section
\ref{sec:model_epjb} divert on average from the expected average
solution. This indicates how well the parameter $q$ can be estimated from
a given degree distribution.

\subsubsection{Discrete description}
\label{sec:recursive-solutionK_epjb}
\begin{figure}[htb]
  \centering
  \includegraphics[width=0.45\textwidth]{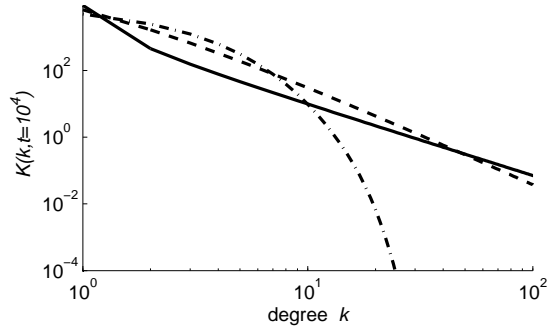}
  \caption{Degree distribution $K(k,t)$ at $t=10^4$ for different values
    of $q$ obtained by recourse of iteration of the discrete equations
    (\ref{eq:level_degree_rec0}-\ref{eq:level_degree_reck}). The
    different lines correspond to: $q=0.0$ (dash-dotted), $q=0.5$
    (dashed) and $q=0.9$ (solid). The plot shows that for increasing
    values of $q$, the distribution is approaches a power law. The
    extreme case $q=1$ corresponds to a star, with the root having
    $k=t-1$ and all other nodes having $k=1$.  }
  \label{fig:degrees_q}
\end{figure}

The evolution of the degree distribution can be formalised as a set of
recursive discrete equations. Let $K(k,t)$ be the {\em number of nodes}
with degree $k$ at time $t$. The initial condition is the following: at
time $t=1$ only one node exists, the root. It has by definition $k=2$
(see equation~(\ref{eq:level_degree_rec0})). 
Equation~(\ref{eq:level_degree_rec0t}) shows that the set of nodes with $k=1$ is decremented by
the expected number of its members being chosen to be linked to a freshly
added node. Furthermore new nodes are added here, hence a one is added.
Finally, the number of nodes with degree $k$ bigger than one are
incremented by the expected number of nodes with degree $k-1$ attracting
a connection to a new node and decremented by the expected number of
nodes with degree $k$ attracting one
(cf.~equation~(\ref{eq:level_degree_reck})). Thus, the whole set of
equations is:
\begin{eqnarray}
  K(k,1)&=& \delta_{k,2}, \label{eq:level_degree_rec0}\\
  K(1,t)&=& K(1,t\!-\!1) +1 -
  (1-q)\frac{K(k,t\!-\!1)}{t}\label{eq:level_degree_rec0t},\\
  K(k,t)&=& K(k,t\!-\!1) \label{eq:level_degree_reck} \\
  & & +(1-q)\frac{K(k\!-\!1,t\!-\!1)-K(k,t\!-\!1)}{t} \nonumber \\
  & &
  +q\frac{(k\!-\!1)K(k\!-\!1,t\!-\!1)-(k\!-\!2)K(k,t\!-\!1)}{t}.\nonumber
\end{eqnarray}
Figure~\ref{fig:degrees_q} shows the numerical solution of these
equations for different values of $q$. First, for $q=0.0$, it can be seen
that the degree distribution is exponential. This is because for this
value, the growth of the tree is equivalent to a fully random
network. For larger values of $q$, the preferential attachment term has
an increasing weight. The curves for $q=0.5$ and $q=0.9$ show that
asymptotically (i.e.~for large values of $k$) the distribution approaches
a scale-free behaviour. The limit case $q=1$, however evolves into a star
as nodes with degree $k=1$ can never be chosen as target of a new
node. Thus, the degree distribution for this case is simply: $K(t-1,t)=
1$, $K(1,t)= t-1$. This fact causes the dent in fgure~\ref{fig:degrees_q}
for $q=0.9$ at $k=1$.

\subsubsection{Closed forms}

\label{sec:closed-formsK_epjb}

As pointed out in Ref~\cite{klemm05}, the model constitutes a particular
case of the network growth model developed in Ref.~\cite{dorogovtsev00},
given that only one link is added each time step. The authors of
Ref.~\cite{dorogovtsev00} also derived a closed form for the stationary
degree distribution in the limit in infinitely large networks (i.e.~when
$t \to \infty$). From this, we can infer the time dependent degree
distribution. We substitute the variables used in
Ref.~\cite{dorogovtsev00} by the ones used in Ref~\citep{klemm05} as
follows: $m=1$ and $a=1+1/q$. The result is
\begin{equation}
  \label{eq:doroEqu9}
  \frac{K(k,t)}{t}=\frac 1 q \frac{\Gamma (2 q^{-1}-1)}{\Gamma
    (q^{-1}-1)}\frac{\Gamma
    (k-1+q^{-1})}{\Gamma (k+2q^{-1})}.
\end{equation}
We use $\Gamma$ to denote the Gamma function. 
For large values of $k$ the asymptotic limit
of the distribution is
\begin{equation}
  K(k)\propto k^{-(1+q)/q}.\label{eq:limit}
\end{equation}
While solving equations in the limit of infinitely large networks is a
common practice in the field of complex networks, one must be cautious
when dealing with real data. The question is whether or not the systems
is large enough to justify the assumption $N \to \infty$.  For example,
real directory structures analysed contain between $10^2$ and $10^5$
nodes.

\begin{figure}[htb]
  \centering
  \includegraphics[width=0.45\textwidth]{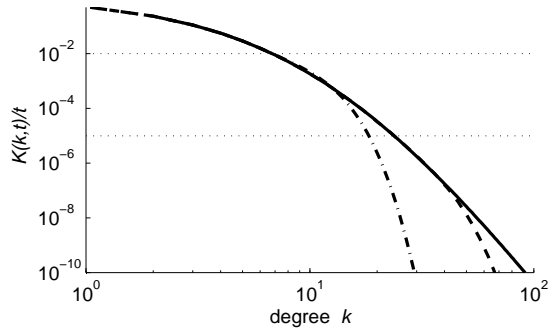}
  \caption{Comparison of the normalised degree distribution $K(k,t)/N$
    for $q=0.1$ for different system sizes with the asymptotic behaviour
    of the degree distribution in the thermodynamic limit
    (cf.~Eq.~\ref{eq:limit}), depicted with solid line. The different
    system sizes are: $N=10^2$ (dash-dotted) and $N=10^5$
    (dashed). Eq.~(\ref{eq:limit}) matches
    Eqs.~(\ref{eq:level_degree_rec0}--\ref{eq:level_degree_reck}) in the
    relevant regions $K(k,t)/N \geq 10^{-2}$ and $K(k,t)/N \geq 10^{-5}$
    (dotted lines).}
  \label{fig:error_limit_q00}
\end{figure}

We have empirically computed the deviation of the numerical solution of
equations~(\ref{eq:level_degree_rec0}--\ref{eq:level_degree_reck}) from
the limit distribution defined by Eq.~\ref{eq:limit}.  The deviation is
strongest for low values of $q$, i.e. $q=0$ is the worst case scenario.
Figure~\ref{fig:error_limit_q00} shows how the thermodynamic limit is
approached for the case $q=0.1$ (Eq.~\ref{eq:limit} is undefined for
$q=0$) for networks of coparable sizes to those found in our data ($10^2$ and $10^5$). The lines $K(k,t)/N=10^{-2}$ and
$K(k,t)/N=10^{-5}$ are marked to indicate the areas relevant for
estimating $q$ for these trees.

To test whether Eq.~\ref{eq:limit} is a sufficient approximation, there
must not be a deviation between Eq.~\ref{eq:limit} and the
Eqs.~(\ref{eq:level_degree_rec0}--\ref{eq:level_degree_reck}) at values
larger than $t^{-1}$.  It can be seen that the degree distribution found
for small system sizes are such that the limit case is still a good
approximation of the distribution found for real systems: The curve for
the system size $N=10^2$ matches Eq.~\ref{eq:limit} for $K(k,t)/N \geq
10^{-2}$. Also for $N=10^5$ the
limiting case is a good approximation for $K(k,t)/N \geq 10^{-5}$. Effectively, the finite-size effects are only
observed with low probability and are all below the $K(k,t) = 1$ line.
For this reason, equation~(\ref{eq:doroEqu9}) could constitute an
appropriate basis for estimating $q$ in a real data set.

\subsubsection{Estimation of $q$ from the degree distribution}
\label{sec:fitting-q-from_epjb}

\begin{figure}[thb]
  \centering
  \includegraphics[width=0.45\textwidth]{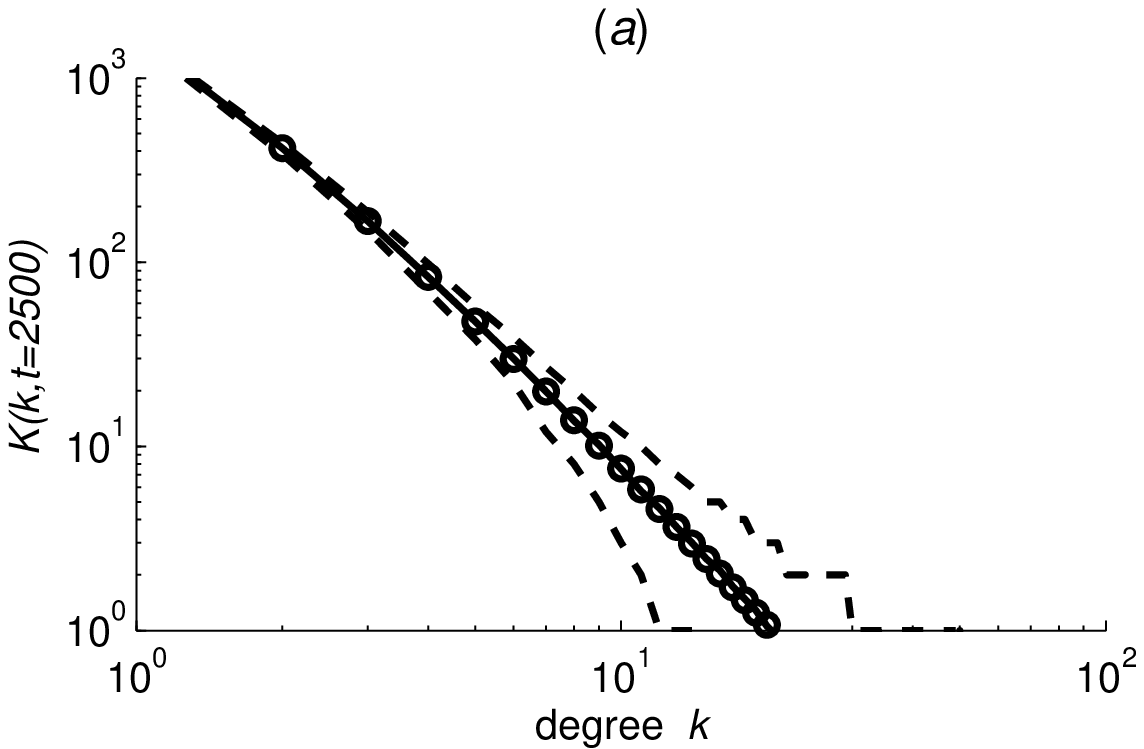}\\
  \includegraphics[width=0.45\textwidth]{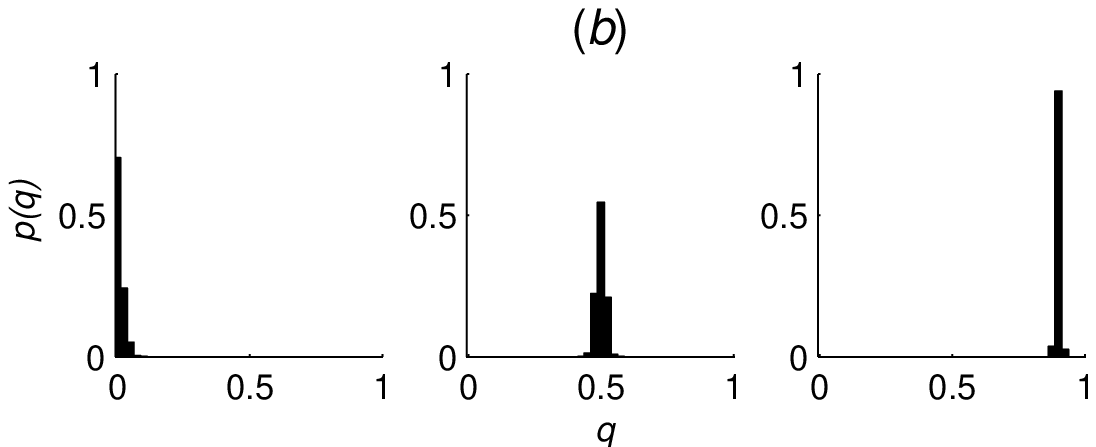}
  \caption{(a) Deviation of single simulated trees from the calculated
    degree distribution ($q=0.5$, $t=2\,500$). The solid line shows the mean
    of the simulations, circles the calculated mean, the dashed lines
    mark the tunnel in which 90\% of the simulated trees lie. Panel (b):
    distribution of estimated values of $q$ by means of the degree
    distribution (see in-line text for details) for trees generated by
    computer simulations of the stochastic model described in Section
    \ref{sec:model_epjb}. In the different plots: $q=0.0$ (left), $q=0.5$
    (middle) and $q=0.9$ (right). The tree size is $N=2\,500$ and the
    distribution is based on $10^4$ simulation runs each.}
  \label{fig:levels_tunnelk}
\end{figure}

When fitting $q$ from the degree distribution of a single data set, it is
important to bear in mind that equation~(\ref{eq:doroEqu9}) only
describes the {\em expected} degree distribution (i.e. the one obtained
after building the average of a large number of concrete tree
manifestations). Particular realisations may deviate from
it. Figure~\ref{fig:levels_tunnelk}(a), shows (with points) the average
value for the degree distribution over $10^3$ realisations of the tree
obtained by numerical simulation. The dashed lines display the intervals
in which 90\% of the degree distributions lie. The expected values
obtained via equation~(\ref{eq:doroEqu9}) are represented with circles.
It can be seen that the intervals around the average values are
relatively narrow.

In order to estimate the value of $q$ for a given tree of size $N$, one
can proceed as follows. First, the degree distribution $K^*(k,t)$ of the
tree, is measured. Then, this distribution is compared to the expected
ones obtained through Eq.~(\ref{eq:doroEqu9}) for different values of
$q$. The value $\bar q_k$ is the one whose associated degree distribution
minimises the root mean square distance to the empirical $K^*(k,t)$.

How accurate the estimation actually is, can be found by determining the
specific error margins while estimating $q$ for a single tree. To do so,
we generated $10^4$ different trees for each $q-$value: $q=0.0$, $q=0.5$
and $q=0.9$, and a system size $N=2\,500$. For each run, $q$ was estimated
by fitting equation~(\ref{eq:doroEqu9}) with the least squares method
described above.  Figure~\ref{fig:levels_tunnelk}(b) shows the distributions
of the estimated values of $q$. In the case of $q=0.5$ the empirically
estimated error margins for $q$ are $[0.48,0.53]$. Then, for $q=0.0$, the
corresponding estimated error margins for $q$ are $[0.0,0.05]$.  For
$q=1.0$ the estimation is always exact as the only possible manifestation
corresponds to a star. For this reason we analysed the case $q=0.9$ and
found error margins of $[0.91,0.89]$.  In all the cases, we set a
confidence level of 90\%. We can conclude that, using this method, the
parameter $q$ can be well approximated by means of the degree
distribution.

\subsection{Level distribution of nodes}
\label{sec:level-distr-nodes_epjb}

At difference with what occurs in non-hierarchical networks, trees
possess a special node, {\em root}, from which the tree starts its
growth. Knowing the dynamics of the distribution of distances towards the
root, unveils an alternative description of the process of tree
growth. In this section, the evolution over time of this level
distribution is solved.  Moreover, it is shown that the equations
describing the growth in terms of the {\em level distribution} are quite
simple for the considered model, and allow for an independent estimation
of the parameter $q$.

\begin{figure}
  \centering
  \includegraphics[width=0.25\textwidth]{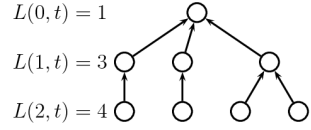}
  \caption{Representation of the tree structure in terms of the level
    distribution: At level 0, there is only one node, the {\em
      root}. From it, the tree is grown with the stochastic model
    described in the text. The level distribution $L(l,t)$ is simply
    given by the number of nodes at a distance $l$ of the root. In the
    figure we represent each link with an arrow from child to parent.}
  \label{fig:leveldist}
\end{figure}

Let $L(l,t)$ be the number of nodes at distance $l$ to the root node at
time $t$; i.e.~$l$ defines the {\em level} of the node. From the set of
the levels of all nodes, the level distribution of the tree can be
compiled (for an illustration see figure \ref{fig:leveldist}).

\subsubsection{Discrete description}
\label{sec:recursive-solutionL_epjb}

In analogy to the recursive description of the degree distribution in
section~\ref{sec:recursive-solutionK_epjb}, we forumulate recursive
equations for the level distribution:
\begin{eqnarray}
  L(l,1) &=& \delta_{l,0} \label{eq:tree_level_reck0}\\
  L(0,t)&=& 1 \label{eq:tree_level_rec0t}\\
  L(l,t)&=& L(l,t\!-\!1) \label{eq:tree_level_reclt} \\
  & &
  +(1-q)\frac{L(l\!-\!1,t\!-\!1)}{t}+q\frac{L(l,t\!-\!1)}{t}, l\geq 1.\nonumber
\end{eqnarray}
First, equation (\ref{eq:tree_level_reck0}) refers to the initial
condition of system in which only one node exists at level zero. Equation
(\ref{eq:tree_level_rec0t}) explicits the condition of uniqueness of the
root node over time. 
To understand Eq.~(\ref{eq:tree_level_reclt}) keep in mind that, adding a
node at level $l$ means that a node at level $l-1$ was selected as parent.
The first non-trivial term -- the one preceded by the factor
$(1-q)$ -- corresponds to the process of random attachment. When nodes
are selected at random, this term is proportional to $L(l-1,t)$. The last
term represents the preferential attachment part, which occurs with
probability $q$. To explain it, one has to consider that the probability to
attach a new node to an existing one in level $l-1$ is proportional to
the sum of the in-degrees on level $l-1$. Interestingly, in a tree, the sum of the
in-degrees on level $l-1$ is equal to the number of nodes in the next
level, i.e.~$L(l,t)$.

\begin{figure}[htb]
  \centering
  \includegraphics[width=0.45\textwidth]{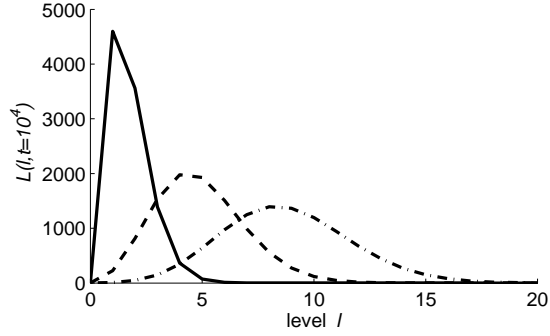}
  \caption{Level distribution $L(l,t)$ at $t=10^4$ for different values
    of $q$ obtained by recourse of iteration of the discrete equations
    (\ref{eq:tree_level_reck0}-\ref{eq:tree_level_reclt}). The different
    lines correspond to: $q=0.0$ (dash-dotted), $q=0.5$ (dashed) and
    $q=0.9$ (solid). The plot shows that for increasing values of $q$,
    the distribution is sharper, corresponding to flatter structures and
    the average level approaches $l=1$. The extreme case $q=1$
    corresponds to a star, with the root node as centre.}
  \label{fig:levels_q}
\end{figure}

Figure \ref{fig:levels_q} shows the expected level distributions obtained
by direct integration of
Eqs.~(\ref{eq:tree_level_reck0}--\ref{eq:tree_level_reclt}) for different
values of $q$ at time $t=10^4$. By increasing $q$, the distribution
shifts closer to the root, and the tree is more shallow. In the limiting
case of $q=1$, the tree takes the form of a star with the root at level
zero and all the other nodes at level 1. Lower values of $q$ produce a
broader level distribution, generating deeper trees. The influence of
time (not shown in the figure) is straight-forward: The larger a tree
grows, the higher the average node depth will be. This effect is stronger
for lower values of $q$. In the next section we investigate the closed
forms description of this relationship.

\subsubsection{Closed forms}
\label{sec:closed-formsL_epjb}

In order to take a closer look at the influence of $t$ on the level
distribution, it is needed to solve the set of
Eqs.~(\ref{eq:tree_level_reck0}-\ref{eq:tree_level_reclt}), which define
its evolution. In particular it is possible to derive closed forms for
the extreme cases $q=1$ and $q=0$.

First, the case of $q=1$ is trivial: it produces a star with the root
node as centre and the $N-1$ other nodes located at level 1, i.e.
\begin{eqnarray}
  L(0,t) = 1 \,;\,\,\,  L(1,t) = t-1.
\end{eqnarray}
The average level $\langle L(l,t) \rangle = 1 - 1/t $ in this case,
approaches the constant value 1 for large enough trees.

Second, by rewriting the discrete time $t$ into the continuous limit, the
following differential equation represents the case $q=0$:
\begin{equation}
  \frac{dL (l,t)}{dt} =  \frac{L(l-1,t)}{t}.
\end{equation}
The initial condition is $L(0,1) = \delta_{1,l}$.
As $L(l,t)$ does not appear on the right hand side
of the differential equation the solution for level $l$ can trivially be obtained by direct integration of the solution for level
$l-1$, divided by $t$. The general solution is found to be
\begin{eqnarray}
  L(l,t) &=& \int_1^t\frac{L(l-1,\tau)}{\tau}d\tau=
  \frac{1}{l!} \left[ \ln(t) \right]^{l}.\label{q0recursion_closed}
\end{eqnarray}
It is easy to see that, in order to obtain the normalised distribution,
the normalisation constant is $N$, i.e. the number of nodes at time
$t$. For any given time, the distribution corresponds to a Poisson
distribution, with parameter $\ln (t)$. Thus, the average level for the
distribution is $\langle L(l,t) \rangle = \ln (t)$ and the variance
$\hbox{Var}( L(l,t) ) = \ln (t)$.

Thus, the broadest level distribution generated by this model has a mean
that grows logarithmically in time.

\subsubsection{Estimation of $q$ from the level distribution}
\label{sec:fitting-q-from-L_epjb}

\begin{figure}[htb]
  \centering
  \includegraphics[width=0.45\textwidth]{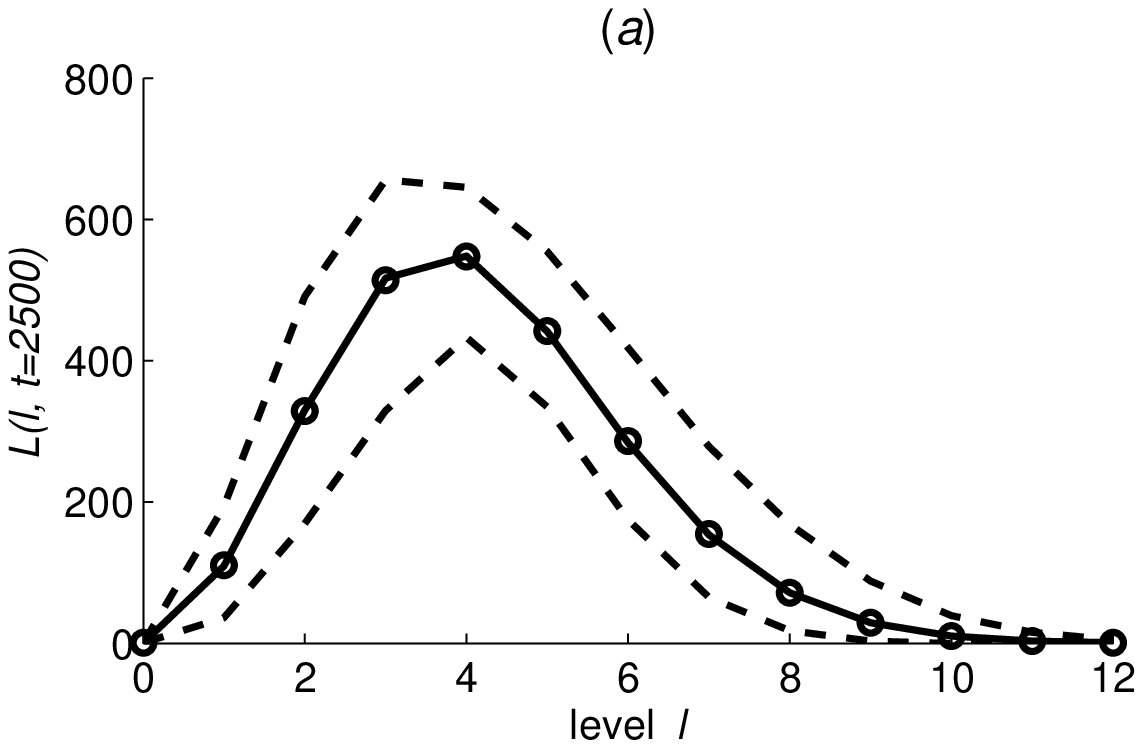}\\
  \includegraphics[width=0.45\textwidth]{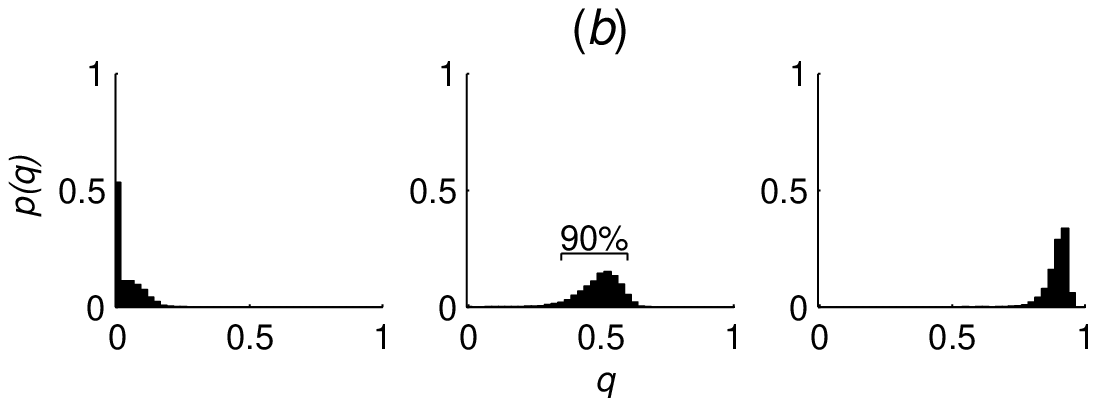}
  \caption{Top: Deviation of single simulated trees from the calculated
    level distribution ($q=0.5$, $t=2\,500$). The solid line shows the mean
    of the simulations, circles the calculated mean, the dashed lines
    mark the tunnel in which 90\% of the simulated trees lie. Bottom:
    Distribution of estimated q for simulations with $q=0.0$ (left),
    $q=0.5$ (middle) and $q=0.9$ (right). The tree size is $N=2\,500$ and
    the distribution is based on $10^4$ simulation runs each.}
  \label{fig:levels_tunnell}
\end{figure}

In a similar fashion as was done for the degree distribution, by means of
equations (\ref{eq:tree_level_reck0}--\ref{eq:tree_level_reclt}) the
expected level distributions can be calculated. Again, the level
distribution obtained from a single realisation of the stochastic model
in Section \ref{sec:model_epjb} might deviate from it. Panel (a) of
figure~\ref{fig:levels_tunnell} shows how large this deviation really
is. For $10^3$ independent trees generated through simulations of the
stochastic model, the dashed lines depict the interval in which 90\% of
the obtained level distributions lie.  The broad intervals for the
expected distribution suggest that estimating the parameter $q$ based on
one tree instance might not be as accurate as the estimation based on the
degree distribution.

Out of an empirically obtained level distribution of a tree with size
$N$, the parameter $q$ is estimated as follows: First, the empirical
level distribution $L^*(l,t)$ is measured. Then, this distribution is
compared to the expected ones
(Eqs.~\ref{eq:tree_level_reck0}--\ref{eq:tree_level_reclt}) obtained for
different values of $q$ and the same integration time $t=N$.  The
estimated value $\bar q_l$ is the one whose associated level distribution
minimises the root mean square distance to the empirical distribution
$L^*(l,t)$.

It is important to know how the deviation of an estimated $q$
from the real value used to synthetically
generate the tree according to the model
(Section~\ref{sec:model_epjb}). This is done in analogy to
the analysis of the estimation based on the degree distribution
(see section~\ref{sec:fitting-q-from_epjb}).  The growth model was
simulated $10^4$ times for three different values of $q$: $q=0.0$,
$q=0.5$ and $q=0.9$ and system size $N=2\,500$.  Then, the value of $q$
was estimated according to the above algorithm.
Figure~\ref{fig:levels_tunnell} (panel (b)) shows the distributions of
the estimated $q$. In the case of $q=0.0$ (left plot), 90\% of the
estimated values are in the interval $[0.0 ,0.13]$. In the case of
$q=0.5$ (middle plot), 90\% of the estimated values lie in the interval
$[0.35,0.60]$, and $q=0.9$ (right plot) yields error margins of
$[0.83,0.95]$.  Finally, for the trees generated with $q=1.0$, the
situation is similar to the one in section~\ref{sec:fitting-q-from_epjb}:
the only possible tree is a star and thus $q$ is always correctly
estimated.

Compared to the accuracy with which $q$ can be calculated from the degree
distribution (see section \ref{sec:fitting-q-from_epjb}) the level
distribution turns out to be a less accurate indicator of
$q$. 

However, it is worth remarking that (if a tree is produced by the process
introduced in Section \ref{sec:model_epjb}) both estimations must agree
quantitatively. In the next section, we test whether this is the case for
user-generated directory structures.

\section{Comparison of real-world data and model}
\label{sec:comparison_epjb}

In the previous theoretical investigations, we have represented the same
model, equation (\ref{eq:model}), in terms of two different
distributions, degree and level distribution. They can be seen as
alternative ways of studying the same tree growth process.  Thus, the two
methods for computing the value of the parameter $q$ can be used to test
whether the growth of a tree occurred following the studied
model. Effectively, if the model is able to correctly reproduce the
degree as well as the level distribution found in real directory
structures, the $q$ calculated based on $L(l,t)$ should strongly
correlate with the $q$ calculated based on $K(k,t)$.

\begin{figure}[htb]
  \centering
  \begin{tabular}{cc}
    \includegraphics[width=0.22\textwidth]{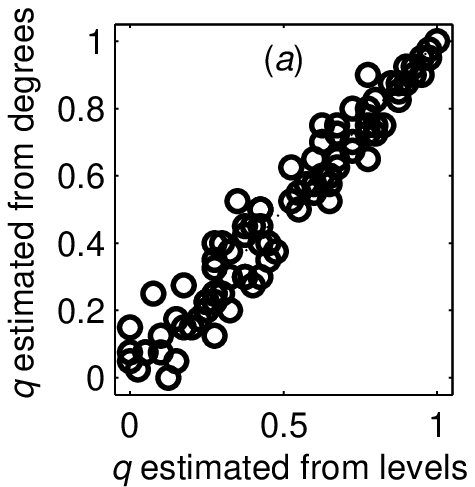}&
    \includegraphics[width=0.22\textwidth]{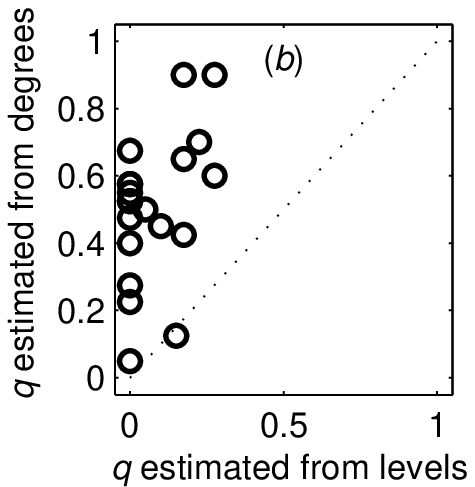}\\
  \end{tabular}
  \label{fig:correlation}
  \caption{In both plots, we compare the values of $q$ estimated through
    the fitting of the level distribution (on the horizontal axis), with
    the estimation of $q$ obtained by means of the degree distribution
    (in the vertical axis).  Panel (a) shows such a comparison for trees
    obtained by direct simulation of the stochastic growth model
    introduced in Section \ref{sec:model_epjb}. The plot consists of
    $100$ trees with values of $q$ in the range $[0,1]$ and size $10^3$.
    In this plot it is apparent a good agreement between the two
    independent measures of the parameter $q$. Panel (b) shows the
    results obtained when analysing $20$ real directory structures with
    sizes between $N=119$ and $N=75\,307$. Interestingly, in this case
    the correlation between the measures is lost. }
\end{figure}

In Figure \ref{fig:correlation}(a), the estimation of $q$ based on the
level distribution (horizontal axis) and degree distribution (vertical
axis) is shown for 100 trees generated by the model of Section
\ref{sec:model_epjb}. Each point corresponds to a tree of size $N=10^3$
and a value of $q$ uniformly drawn within the interval $[0,1]$. As
expected, both measures are strongly correlated, and the points barely
depart from the identity function.

In order to test whether the same applies to directory structures, we
have collected $20$ user-generated directories corresponding to
Linux/UNIX computer facilities.  We have only considered directories as
nodes of the network and did not include files or hard or soft links in
the network. Also configuration directories (those with a leading dot,
which are automatically generated either by the system or by particular
programs) have been discarded, as they are not consciously generated by
the user, and they present approximately the same structure for every
user. With this, the trees obtained contain between $N=119$ and
$N=75\,307$ directories (with median: $3467$).

Figure~\ref{fig:correlation}(b) shows correlation between the two
estimation methods when applied to the directory data collected. It can
be seen that the correlation between the two measures is lost. Thus, the
two estimated values of $q$ are incompatible with each other. This leads
us to the conclusion that the model by Klemm {\em et al.} in its current
state reproduces the degree distributions of directory structures quite
well as shown in Ref.~\cite{klemm05}, but fails to produce the
corresponding level distribution.

It is interesting to note that the parameter $q$ is wide spread, covering
the range $[0,0.9]$ when estimating it by means of the degree
distribution. This is in agreement with the experimental findings of
Ref.~\cite{klemm05}, although in that reference an alternative Monte
Carlo method was applied. However, when the level distribution is used to
estimate the parameter $q$, the values found lay in the interval
$[0,0.28]$. This implies that the tree structures found in real-world
directory structures are much deeper than the predicted by the model.

It could be argued that the values of $q$ measured are lower because
users might start their directory structure after a phony directory, such
as the {\tt Desktop} folder. Yet, performing the same regression analysis
on shifted level distributions shows that in most of the cases the
distribution must be shifted 3 or more levels in order to improve the
correlation between both estimators of $q$. Such shifts, it is important
to remark, are unrealistic in this context.

\section{Conclusions}
\label{sec:conclusion_epjb}

In this paper we have investigated a stochastic growth model for trees,
where a parameter $q$ interpolates between two limiting cases: random
growth ($q=0$) and preferential attachment ($q=1$). This model has been
previously used to model the evolution of user-generated directories
\cite{klemm05,klemm06}, in particular the properties of the degree
distribution and allometric scaling.

In this paper we extend the current state of this research by means of
three contributions: 

(i) We propose an alternative way of estimating the parameter $q$ from
data by fitting an analytical solution.  We show that, even though finite
size effects exist, the solution proposed in Ref~\cite{dorogovtsev00} for
the thermodynamic limit is sufficiently accurate to estimate $q$
analytically from the data.  This approach is more efficient than the
computation intensive approach used in Ref~\cite{klemm05}.

(ii) We introduce the concept of level distribution as an important
characterisation of trees. We argue that in order to verify a tree growth
model, in addition to the degree distribution also the level distribution
has to be taken into account. A model can claim evidence only if both of
these independent representations are matched by the data. In the
particular case of the stochastic growth model described above, it means
that both ways should lead to the same estimation of the parameter $q$.

(iii) Applying our results for the degree and the level distribution to
both simulated and user generated data, we find a perfect correlation
between the estimated $q$ values for the simulated trees, but no
correlation for the real-world user generated directories. Hence, we have
to conclude that the growth of real directory trees are not sufficiently
captured by the model given in \cite{klemm05}. In particular, user
directories extend more in depth than the model predicts.

Our contributions also highlight that an analysis proven to be of
relevance for complex networks does not necessarily give the full
description of hierarchical structures, be they real or simulated. Thus,
different aspects (or complementary descriptions, as was done in this
Paper) must be studied in order to fully characterise these structures.

{\bf Acknowledgement:} We want to thank the anonymous users who run our
script to provide us with data on their directory structures. CJT
acknowledges financial support from SBF (Swiss Confederation) through
research project C05.0148 (Physics of Risk).

\bibliographystyle{abbrv}
\bibliography{index}

\begin{thebibliography}{23}

\bibitem{caldarelli05}
G.~Caldarelli, \emph{Scale--Free Networks} (Oxford University Press, 2007)

\bibitem{rodrigueziturbe97}
I.~Rodr{\'\i}guez-Iturbe, A.~Rinaldo, \emph{{Fractal River Basins: Chance and
  Self-Organization}} (Cambridge University Press, 1997)

\bibitem{weibel91}
E.~Weibel, American Journal of Physiology- Lung Cellular and Molecular
  Physiology \textbf{261}(6), 361 (1991)

\bibitem{zamir99}
M.~Zamir, Journal of Theoretical Biology \textbf{197}(4), 517 (1999)

\bibitem{banavar99}
J.R. Banavar, A.~Maritan, A.~Rinaldo, Nature \textbf{399}(6732), 130 (1999)

\bibitem{prusinkiewicz90}
P.~Prusinkiewicz, A.~Lindenmayer, \emph{{The algorithmic beauty of plants}}
  (Springer-Verlag New York, Inc. New York, NY, USA, 1990)

\bibitem{cracraft04}
J.~Cracraft, M.~Donoghue, \emph{{Assembling the Tree of Life}} (Oxford
  University Press, USA, 2004)

\bibitem{dupuis84}
C.~Dupuis, Annual Reviews in Ecology and Systematics \textbf{15}(1), 1 (1984)

\bibitem{rokas06}
A.~Rokas, Science \textbf{313}, 1897 (2006)

\bibitem{tessone07c}
E.A. Herrada, C.J. Tessone, V.M. Egu\'{\i}luz, E.~Hern\'andez-Garc\'ia, C.M.
  Duarte, PLoS ONE \textbf{3}, e2757 (2008)

\bibitem{hernandez-garcia07}
E.~Hern{\'a}ndez-Garc{\'i}a, E.A. Herrada, A.F. Rozenfeld, C.J. Tessone, V.M.
  Egu{\'i}luz, C.M. Duarte, S.~Arnaud-Haond, E.~Serr{\~a}o, \emph{Evolutionary
  and Ecological Trees and Networks}, in \emph{XV Conference on Nonequilibrium
  Statistical Mechanics and Nonlinear Physics}, edited by O.~Descalzi, O.A.
  Rosso, H.A. Larrondo (AIP Conference Proceedings, 2007), Vol. 913, pp. 78--83

\bibitem{muchnik07}
L.~Muchnik, R.~Itzhack, S.~Solomon, Y.~Louzoun, Physical Review E
  \textbf{76}(1), 016106 (~12) (2007)

\bibitem{huffman52}
D.A. Huffman, Proceedings of the IRE \textbf{40}(9), 1098 (1952)

\bibitem{knuth97}
D.~Knuth, \emph{{The art of computer programming, volume 1: fundamental
  algorithms}} (Addison Wesley Longman Publishing Co., Inc. Redwood City, CA,
  USA, 1997)

\bibitem{goodrich02}
M.~Goodrich, R.~Tamassia, \emph{{Algorithm Design: Foundations, Analysis, and
  Internet Examples}} (J. Wiley, 2002)

\bibitem{golder05}
S.~Golder, B.~Huberman, Journal of Information Science \textbf{32}, 198 (2006)

\bibitem{codd70}
E.~Codd, Communications of the ACM \textbf{13}(6), 377 (1970)

\bibitem{klemm05}
K.~Klemm, V.M. Egu\'{\i}luz, M.~San~Miguel, Physical Review Letters
  \textbf{95}(12), 128701 (~4) (2005)

\bibitem{klemm06}
K.~Klemm, V.M. Egu{\'\i}luz, M.S. Miguel, Physica D: Nonlinear Phenomena
  \textbf{224}(1-2), 149 (2006)

\bibitem{krapivsky00}
P.L. Krapivsky, S.~Redner, F.~Leyvraz, Physical Review Letters \textbf{85}(21),
  4629 (2000)

\bibitem{garlaschelli03}
D.~Garlaschelli, G.~Caldarelli, L.~Pietronero, Nature \textbf{423}, 165 (2003)

\bibitem{garlaschelli05}
D.~Garlaschelli, G.~Caldarelli, L.~Pietronero, Nature \textbf{E4}, 165 (2005)

\bibitem{dorogovtsev00}
S.~Dorogovtsev, J.~Mendes, A.~Samukhin, Physical Review Letters
  \textbf{85}(21), 4633 (2000)

\end{thebibliography}

\end{document}